# Mixed-Salt Enhanced Chemical Vapor Deposition of Two-Dimensional Transition Metal Dichalcogenides

*Shisheng Li,*[1] *Yung-Chang Lin,*[2] *Jinhua Hong,*[2] *Bo Gao,*[3,4] *Hong En Lim,*[5] *Xu Yang,*[6] *Song Liu,*[7] *Yoshitaka Tateyama,*[3,4] *Kazuhito Tsukagoshi,*[3] *Yoshiki Sakuma,*[6] *Kazu Suenaga,*[2,8] *and Takaaki Taniguchi*[3]

1 International Center for Young Scientists (ICYS), National Institute for Materials Science (NIMS), Tsukuba 305-0044, Japan

2 Nanomaterials Research Institute, National Institute of Advanced Industrial Science and Technology, AIST Central 5, Tsukuba 305-8565, Japan

3 International Center for Materials Nanoarchitectonics (WPI-MANA), National Institute for Materials Science (NIMS), Tsukuba 305-0044, Japan

4 Center for Green Research on Energy and Environmental Materials (GREEN), National Institute for Materials Science (NIMS), Tsukuba 305-0044, Japan

5 Department of Physics, Tokyo Metropolitan University, Hachioji 192-0397, Japan

6 Research Center for Functional Materials, National Institute for Materials Science (NIMS), Tsukuba 305-0044, Japan

7 Institute of Chemical Biology and Nanomedicine (ICBN), College of Chemistry and Chemical Engineering, Hunan University, Changsha 410082, P. R. China

8 The Institute of Scientific and Industrial Research (ISIR-SANKEN), Osaka University, Osaka 567-0047, Japan

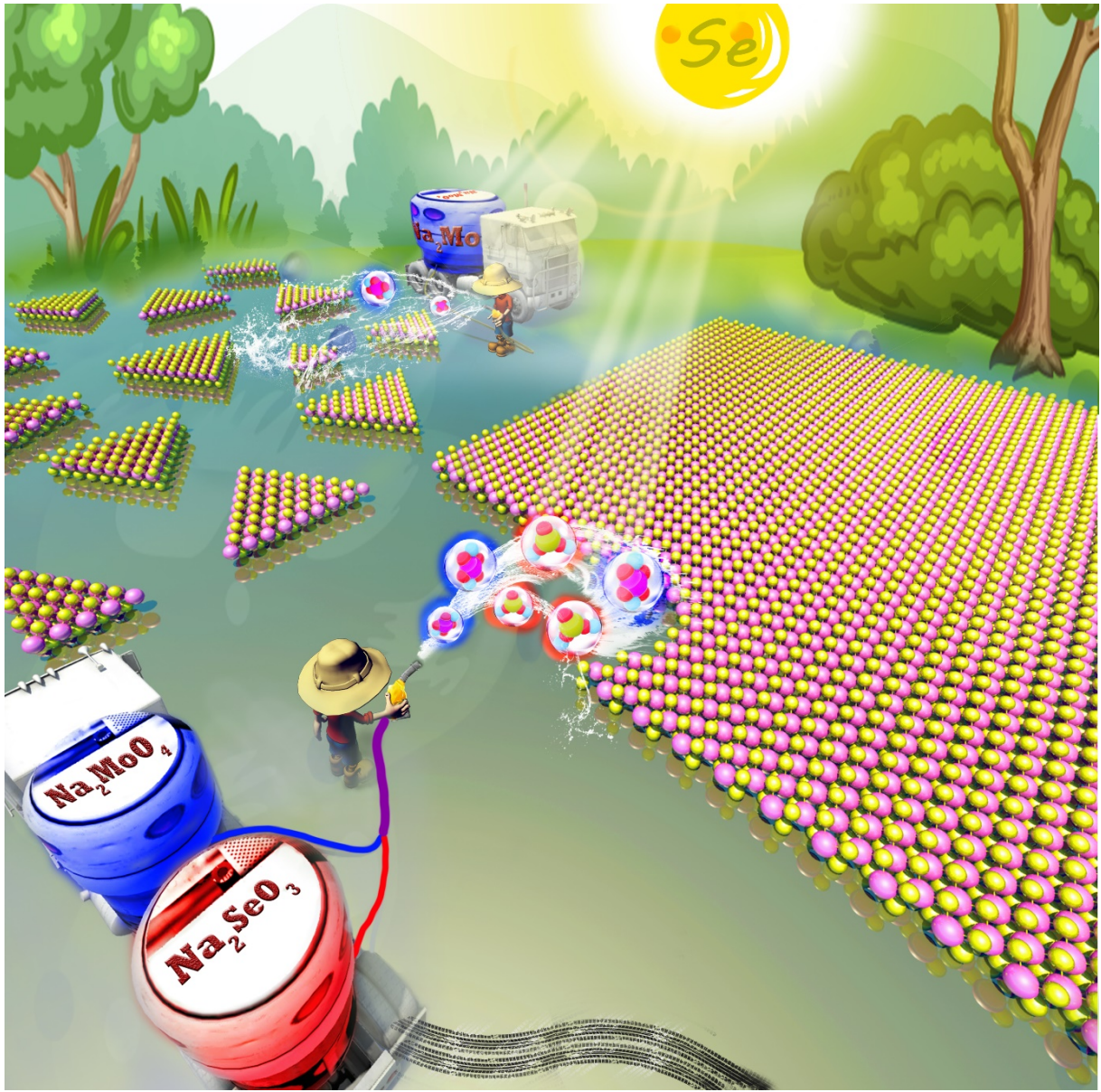

**Boosting growth of MoSe$_2$ with Na$_2$SeO$_3$.** An enhanced CVD growth of continuous MoSe$_2$ film was achieved by adding Na$_2$SeO$_3$ to Na$_2$MoO$_4$. This work provides a general strategy for large-area growth of 2D transition metal dichalcogenides using mixed molten salts.

**Abstract**

The usage of molten salts, e.g., Na$_2$MoO$_4$ and Na$_2$WO$_4$, has shown great success in the growth of two-dimensional (2D) transition metal dichalcogenides (TMDCs) by chemical vapor deposition (CVD). In comparison with the halide salt (i.e., NaCl, NaBr, KI)-assisted growth (Salt 1.0), the molten salt-assisted vapor-liquid-solid (VLS) growth technique (Salt 2.0) has improved the reproducibility, efficiency and scalability of synthesizing 2D TMDCs. However, the growth of large-area MoSe$_2$ and WTe$_2$ is still quite challenging with the use Salt 2.0 technique. In this study, a renewed Salt 2.0 technique using mixed salts (e.g., Na$_2$MoO$_4$-Na$_2$SeO$_3$ and Na$_2$WO$_4$-Na$_2$TeO$_3$) is developed for the enhanced CVD growth of 2D MoSe$_2$ and WTe$_2$ crystals with large grain size and yield. Continuous monolayer MoSe$_2$ film with grain size of 100-250 μm or isolated flakes up to ~ 450 μm is grown on a halved 2-inch SiO$_2$/Si wafer. Our study further confirms the synergistic effect of Na$^+$ and SeO$_3^{2-}$ in the enhanced CVD growth of wafer-scale monolayer MoSe$_2$ film. And thus, the addition of Na$_2$SeO$_3$ and Na$_2$TeO$_3$ into the transition metal salts could be a general strategy for the enhanced CVD growth of many other 2D selenides and tellurides.

**INTRODUCTION**

Two-dimensional (2D) layered transition-metal dichalcogenides (TMDCs) with unique electrical properties are promising materials for future electronic devices. Among them, the semiconducting TMDCs (e.g., MoS$_2$, MoSe$_2$, WSe$_2$, etc.) are idea channel materials for field effect transistors (FETs) because of their high carrier mobility and current on/off ratio, which may reach from tens to hundreds cm$^2$/Vs and up to 10$^8$, respectively.[1-3] Meanwhile, the metallic

TMDCs (e.g., VSe$_2$, VTe$_2$, WTe$_2$, etc.)[4-7] also play important roles in future all-2D integrated circuits as contacts and interconnects.[8] To realize their potential application in 2D electronics, the TMDCs should be prepared with high crystallinity, uniformity and reproducibility at low cost.

In the past decade, many growth techniques were employed for the synthesis of 2D TMDCs, from chemical vapor transport (CVT) growth of bulk TMDC crystals to the pioneering chemical vapor deposition (CVD) growth of MoS$_2$, and then metal organic (MO) CVD synthesis of large-area 2D TMDCs.[9-12] Among these approaches, CVD is the most promising method due to its low cost, easy controllability, and scalability for mass production. The typical CVD growth of 2D TMDCs usually involves the powders of transition metal oxides and chalcogen precursors, where the 2D TMDCs are grown via a vapor-solid-solid (VSS) growth mechanism. However, some oxides (e.g., WO$_3$ and Nb$_2$O$_5$) with high melting points close to 1500 °C are hard to vaporize at the common growth temperature of 700-900 °C.[13] To solve this issue, we first introduced halides (e.g., NaCl, NaBr, KI, etc.) as growth promoters to facilitate the growth of large WS$_2$ and WSe$_2$ monolayers.[14] Such improved CVD growth is attributed to the in-situ generated oxyhalides, which have low melting points around 300 °C and thus providing sufficient high vapor pressure for the feasible growth of 2D TMDCs. This Salt 1.0 technique had succeeded in growing tens different kinds of layered metal chalcogenides and their alloys/junctions.[15] A recent study found that the fluoride can also facilitate the growth of graphene and h-BN monolayers.[16] However, the Salt 1.0 technique following the VSS growth mechanism has several intrinsic drawbacks, which include non-uniform metal precursor (vapor) distribution on the growth substrate, poor coverage uniformity and reproducibility, and contamination on growth chambers. These cause the wafer-scale and patterned growth of TMDCs almost impossible. Furthermore, extra costs are needed to replace the contaminated growth chambers/tubes frequently.

Until recently, Salt 2.0 technique has been developed to enable the CVD growth of 1D and 2D TMDCs via a vapor-liquid-solid (VLS) mechanism.[17-20] This method uses the molten salts, e.g., $Na_2Mo_2O_7$, $Na_2MoO_4$, $Na_2WO_4$, $NaReO_4$, $NaVO_3$, etc., which have melting points close to the typical TMDC growth temperatures (650-850 °C). During the reaction, these transition metal salts are in molten liquid state, which is essentially different from those in the Salt 1.0 technique. The molten salts have the advantage to be prepared in stable aqueous solutions that can either be uniformly deposited on large wafers via spin-coating or to form patterns on substrates. We had indeed demonstrated 2-inch wafer-scale and deterministic patterned growth of $MoS_2$ and $WS_2$ monolayers using $Na_2MoO_4$ and $Na_2WO_4$ aqueous solutions, respectively.[18, 19] Zuo et al., also demonstrated the growth of $MoS_2$, $MoSe_2$, $WS_2$ and $WSe_2$ inside the channels of optical fibers using the Salt 2.0 technique, resulting in an ultrahigh nonlinearity ever achieved.[21] In addition, we also demonstrated the CVD growth of rhenium (Re) and vanadium (V)-doped TMDCs by simply mixing molten salt aqueous solutions in designated ratios. These CVD-grown Re- and V-doped TMDCs with tunable dopants concentrations, electrical and optical properties are highly promising for future electronic and optoelectronics.[22] The Salt 2.0 technique is hence, highly promising towards the practical application of 2D TMDC-based electronics.

In this paper, we report an enhanced CVD growth of 2D $MoSe_2$, $ReSe_2$ and $WTe_2$, which are still quite challenging to grow with the Salt 2.0 technique using $Na_2MoO_4$, $NaReO_4$, and Na$_2$WO$_4$ only. By using mixed transition metal salts and halogen salts, e.g., Na$_2$MoO$_4$-Na$_2$SeO$_3$, NaReO$_4$-Na$_2$SeO$_3$ and Na$_2$WO$_4$-Na$_2$TeO$_3$, much improved growth of MoSe$_2$, ReSe$_2$ and WTe$_2$ is achieved, respectively. The most substantial achievement lies in the growth of continuous MoSe$_2$ film on a halved 2-inch SiO$_2$/Si wafer with large grain sizes of 100-250 μm or isolated flakes up to ~ 450 μm using the Na$_2$MoO$_4$-Na$_2$SeO$_3$, which is much better than that grown using Na$_2$MoO$_4$ only. Similarly, the ReSe$_2$ and WTe$_2$ also show much improved yield and grain size with the addition of Na$_2$SeO$_3$ and Na$_2$TeO$_3$, respectively. Our theoretical studies indicate that the growth of 2D Se- and Te-based TMDCs is thermodynamically favored when the chalcogen salts are introduced. The chalcogen salts added can effectively decrease the formation energies of 2D TMDCs. Such enhanced CVD growth using mixed salts strategy therefore provides a robust synthetic alternative for many 2D selenides and tellurides towards future electronics.

**RESULTS AND DISSCUSSIONS**

**Strategy for enhanced CVD growth of 2D TMDCs using mixed salt precursors**

The successful growth of 2D TMDCs with the Salt 2.0 technique requires a careful choice of salt precursors. Basically, the molten salts need to meet two main criteria: 1) possess reasonably high melting points; 2) formation of stable aqueous solutions, considering the growth mechanism involved. Take the VLS growth of 1D MoS$_2$ nanoribbons for instance, molten Na-Mo-O salt (Na$_2$Mo$_2$O$_7$) first forms liquid droplets on the growth substrates. Then, the sulfur vapor adsorbs on their surface and diffuses into the droplets, forming Na-Mo-S-O. When the Na-Mo-O-S droplets become supersaturated with sulfur, solid MoS$_2$ nanoribbons are precipitated out finally.[17] And thus, precursors with a suitable range of heat tolerance/stability and easy incorporation of chalcogen are crucial to initiate seeding and to mediate growth.

Figure 1a shows some transition metal salts that have been used for the successful growth of 2D transition metal sulfide.[17, 21, 23] However, it is still quite challenging to grow large-area, high-quality 2D transition metal selenide (e.g., $MoSe_2$) and tellurides (e.g., $WTe_2$) using these precursors. We attribute this to the difficult incorporation of chalcogen atoms, especially Se and Te in the transition metal salts to initiate the TMDC nucleation and the subsequent growth. In previous study, tellurium was mixed with $MoO_3$, tungsten and rhenium to promote the growth 2D transition metal sulfides, $MoS_2$, $WS_2$ and $ReS_2$, respectively.[24, 25] To overcome this, chalcogen salts, e.g., $Na_2SeO_3$ and $Na_2TeO_3$ with similar chemical properties (reasonably high melting points and soluble in $H_2O$) were added to the transition metal salts to prepare the mixed-salt aqueous solutions. Compared to the molten salts with Na-Mo-O or Na-W-O only, the mixed molten salts containing Na-Mo-Se-O or Na-W-Te-O may assist the nucleation during the temperature ramping process and facilitate the rapid growth of 2D TMDCs when elemental chalcogen vapor is introduced at the growth stage. To confirm this idea, we first prepared the mixed-salt aqueous solutions. Then, the solutions were spin-coated onto the growth substrates. CVD growth was performed in a 2-inch tube furnace with Se or Te powder loaded in the upstream and growth substrates positioned at the center. The synthetic strategy and method are illustrated in Figure 1b and Figure S1.

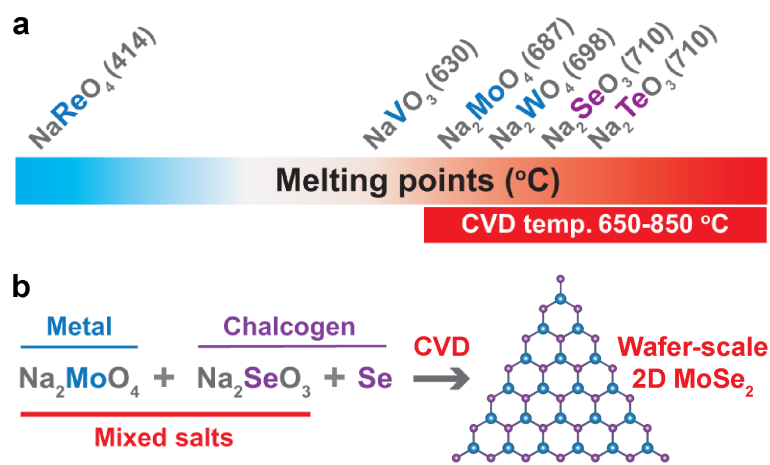

**Figure 1. Enhanced CVD growth of 2D TMDCs with mixed transition metal salts and chalcogen salts.** (a) A color bar indicating the melting points of several salt precursors. (b) The synthetic strategy for enhanced CVD growth of wafer-scale 2D MoSe$_2$ films with mixed-salt precursors.

**Enhanced CVD growth of MoSe$_2$ monolayers using mixed Na$_2$MoO$_4$-Na$_2$SeO$_3$**

To evaluate the proposed strategy for enhanced CVD, we first conducted the growth of MoSe$_2$ monolayers using the mixed Na$_2$MoO$_4$-Na$_2$SeO$_3$ (16 mM, 1:1) solution. Figure 2a is a photo showing a uniform MoSe$_2$ film grown on a halved 2-inch SiO$_2$/Si wafer, which has an obvious color contrast different from the bare SiO$_2$/Si wafer. Low-magnification optical image shows that the MoSe$_2$ film grown with Na$_2$MoO$_4$-Na$_2$SeO$_3$ has a full coverage with a fairly low-density of thick-layers (Figure 2b). Our statistical result indicates a high coverage yield of MoSe$_2$ monolayer up to ~ 98.3%. A fine control of the mixed salts' concentrations is anticipated to result in a full coverage of MoSe$_2$ monolayer in future. High-magnification optical images reveal that the as-grown MoSe$_2$ monolayer has large domain size in the range of 100-250 μm (Figure 2c, Figure S2a and S2b). When the Na$_2$MoO$_4$:Na$_2$SeO$_3$ molar ratio changes to 1:2, large MoSe$_2$ monolayers with a domain size up to ~450 μm are observed (Figure S2c). Meanwhile in the controlled growth without adding any Na$_2$SeO$_3$, only small (~ 10 μm) MoSe$_2$ flakes are grown (Figure 2d and Figure S2d).

Figure 2e is a typical Raman spectrum of as-grown monolayer MoSe$_2$ film with the characteristic A$_{1g}$ peak located at 239.8 cm$^{-1}$, which is slightly higher than the strain-free exfoliated MoSe$_2$ monolayers, ~239.5 cm$^{-1}$.[26] Figure 2f shows an optical image and corresponding Raman maps of two adjacent monolayer MoSe$_2$ domains. The Raman peak position map over a large area of 165 × 355 μm$^2$ demonstrates a small variation from 238.9 to 239.8 cm$^{-1}$. We attribute the Raman peak variation to the strain existed in the CVD-grown monolayer MoSe$_2$ film.[27, 28] Nonetheless, the Raman peak intensity map shows a good uniformity. The intensity gradient from the bottom left to top right is attributable to the different focusing conditions encountered in such a large mapping area caused by the uneven substrate/sample stage. Figure 2g is a typical photoluminescence (PL) spectrum of as-grown monolayer MoSe$_2$ with a peak position located at ~837 nm (~1.481 eV) corresponding to the A exciton emission. The small FWHM of ~ 40 meV indicates a good optical quality of as-grown monolayer MoSe$_2$ film.[26, 29-31] Figure 2h shows an optical image and corresponding PL maps of two adjacent monolayer MoSe$_2$ domains. In the large mapping area of 120 × 300 μm$^2$, the PL maps demonstrate small deviation in peak position (photon energy) and peak intensity, indicating a great optical uniformity in such large monolayer MoSe$_2$ domains.[29, 30]

The structural analysis performed on the transferred MoSe$_2$ monolayer via annular dark-field (ADF) scanning transmission electron microscopy (STEM) reveals that the MoSe$_2$ monolayer has a good crystallinity (Figure 2i). A small portion of Se vacancies is observed, possibly due to the hydrogen etching occurred during the cooling process. We further investigated the electrical quality of the as-grown MoSe$_2$ monolayers by fabricating FETs (Figure S3a-c). The MoSe$_2$-FETs show typical n-type transport behavior with a high current on/off ratio up to $10^8$ (Figure 2j). From a total 16 devices measured, they show current on/off ratios around $10^8$-$10^9$ and electron mobilities range from 5.2 to 16.4 cm$^2$/Vs with an average of 8.2 cm$^2$/Vs (Figure S3d). It is noteworthy to mention that the electrical performance of our CVD-grown monolayer MoSe$_2$ film is similar or to a certain extent better than the previously reported, CVD-grown isolated flakes and continuous film.[32-37] Although high-mobility monolayer MoSe$_2$ have also been grown by Wang et al. and Li et. al.,[29, 38] they are limited only to isolated submillimetre-size flakes. It is therefore crucial to seek a feasible growth that accommodates large-area growth of high-quality MoSe$_2$ film. Furthermore, the value obtained is still incomparable to that of the CVD-grown MoS$_2$ monolayers,[18, 39] the electrical performance, environmental stability and crystallinity of our CVD-grown monolayer MoSe$_2$ films can be further improved by using a fast-cooling process with high-flux of Se vapor.[40] Notably, enhanced CVD growth of 2D ReSe$_2$ flakes is also achieved via the use of mixed NaReO$_4$-Na$_2$SeO$_3$ (Figure S4). It could be a versatile strategy for growing many other 2D metal selenides.

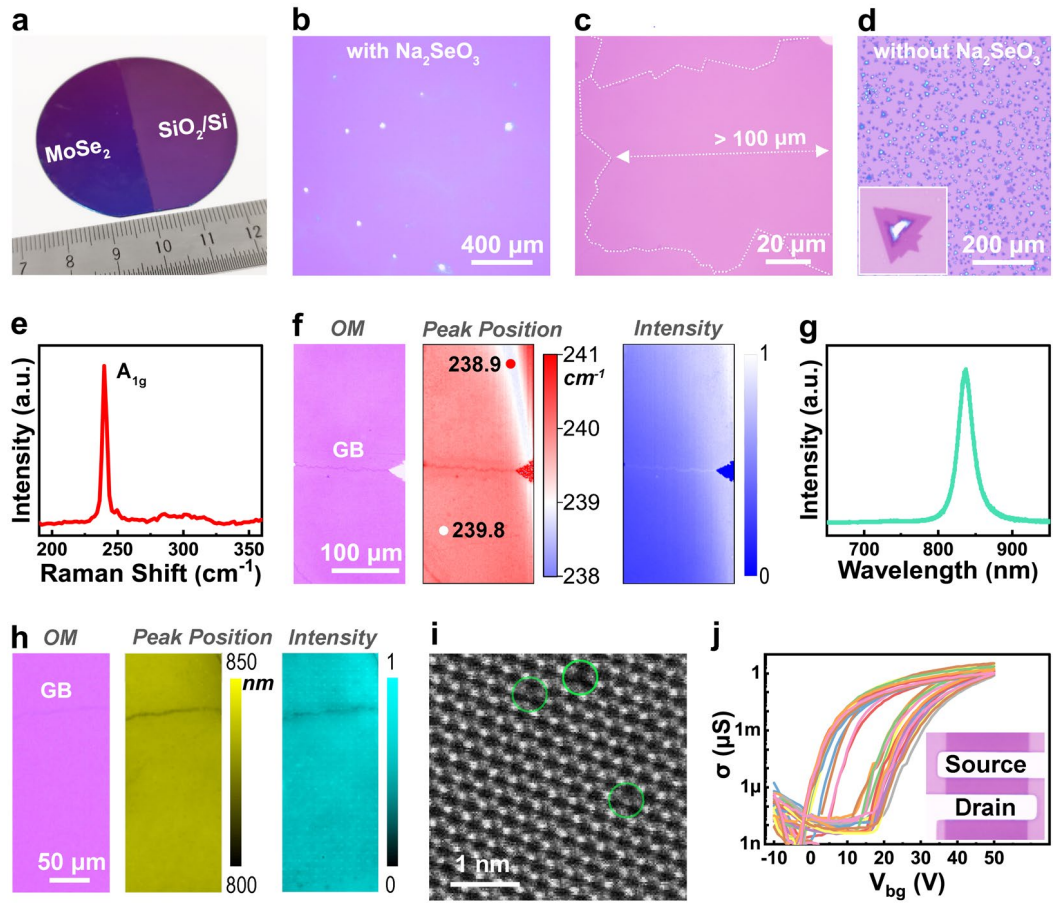

**Figure 2. Characterization of MoSe$_2$ monolayers grown with Na$_2$MoO$_4$-Na$_2$SeO$_3$.** (a) A photo of monolayer MoSe$_2$ film grown on a halved 2-inch wafer. (b, c) Optical images of monolayer MoSe$_2$ film grown with the mixed salts. The dotted lines indicate the grain boundaries (GB). (d) An optical image of MoSe$_2$ grown with Na$_2$MoO$_4$ only. Inset shows a typical MoSe$_2$ flake in an area of 10 ×10 μm$^2$. (e) Typical Raman spectrum of the as-grown MoSe$_2$ monolayer. (f) OM image and corresponding Raman peak position and intensity maps. (g) Typical PL spectrum of the as-grown MoSe$_2$ monolayer. (h) OM image and corresponding PL peak position and intensity maps. (i) A typical ADF-STEM image of the as-grown MoSe$_2$ monolayer. The green circles indicate the single and double Se vacancies in the MoSe$_2$ lattice. (j) Transport curves of MoSe$_2$-FETs. Inset shows the channel area of a MoSe$_2$-FET. The data presented in figures e-j were obtained from monolayer MoSe$_2$ films grown with the mixed salts.

**Enhanced CVD growth of WTe$_2$ using mixed Na$_2$WO$_4$-Na$_2$TeO$_3$**

We further evaluated the CVD growth of WTe$_2$ using mixed Na$_2$WO$_4$-Na$_2$TeO$_3$ and Na$_2$WO$_4$ only. Figure 3a and 3b are optical images of few-layer WTe$_2$ flakes grown on sapphire substrate with Na$_2$WO$_4$-Na$_2$TeO$_3$ (30 mM, 1:1). In contrast to the thick, small flakes grown (Figure 3c and inset), addition of the Na$_2$TeO$_3$ results in the growth of thin WTe$_2$ of much larger size (Figure 3a and 3b). Figure 3d is an atomic force microscope (AFM) image obtained from a WTe$_2$ flake grown with the mixed salts. Most of the WTe$_2$ flakes have a height of ~2.1 nm, corresponding to a tri-layer structure (Figure 3d). A typical Raman spectrum of as-grown WTe$_2$ flake on sapphire substrate is presented in Figure 3e. Five characteristic peaks located at 110.7 cm$^{-1}$, 116.6 cm$^{-1}$, 134.0 cm$^{-1}$, 163.0 cm$^{-1}$, and 211 cm$^{-1}$ are well-matched to the Raman active modes ($A_2^4$, $A_1^3$, $A_1^5$, $A_1^7$, $A_1^9$) of T′-WTe$_2$.[41] Figure 3f shows a typical high-resolution ADF-STEM image and the atomic structure of T′-WTe$_2$. The typical transport curve of a tri-layer WTe$_2$-FET in Figure 3g shows good metallic transport behaviour with poor gate dependency and high conductivity. It could be a promising contact material for future 2D electronics.[6, 7]

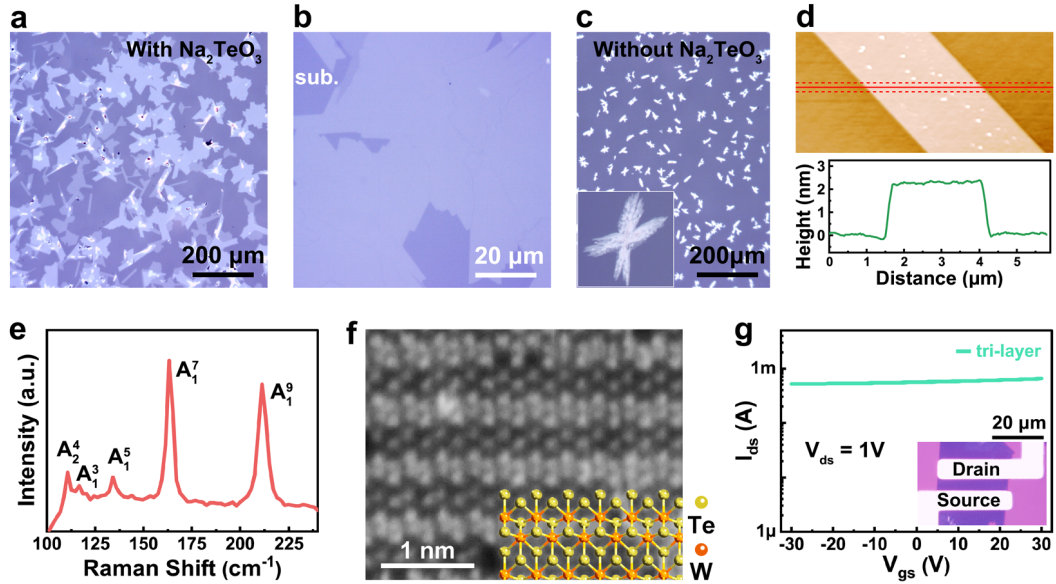

**Figure 3. Characterization of the WTe$_2$ grown with Na$_2$WO$_4$-Na$_2$TeO$_3$.** (a, b) Optical images of few-layer WTe$_2$ grown with the mixed Na$_2$WO$_4$-Na$_2$TeO$_3$. (c) An optical image of the WTe$_2$ grown with Na$_2$WO$_4$ only. Inset shows a single WTe$_2$ flake in an area of 10 ×10 μm$^2$. (d) AFM image and height profile of a WTe$_2$ flake. (e) Raman spectra of as-grown T′-WTe$_2$. (f) Typical ADF-STEM image and atomic models of as-grown T′-WTe$_2$. (g) Typical transport curve of a tri-layer WTe$_2$- FET. Inset shows the optical image of the channel area. Data in figure d-g were obtained from the WTe$_2$ flakes grown with the mixed salts.

**The mechanism for improved CVD growth of 2D TMDCs with mixed salts**

To reveal the mechanism for such enhanced CVD growth using mixed salts, we compared the MoSe$_2$ growth in separate sets of experiments to identify the functions of SeO$_3^{2-}$ and Na$^+$, respectively.

First, we clarify the effectiveness of SeO$_3^{2-}$ by using salt precursors without/with SeO$_3^{2-}$ at fixed Na$^+$ and MoO$_4^{2-}$ molar ratio, i.e. Na$_2$MoO$_4$ (16 mM) and (NH$_4$)$_2$MoO$_4$-Na$_2$SeO$_3$ (16 mM, 1:1). Figure 4a shows a halved 2-inch SiO$_2$/Si wafer fully covered with a uniform MoSe$_2$ film grown with (NH$_4$)$_2$MoO$_4$-Na$_2$SeO$_3$, where Na$^+$:MoO$_4^{2-}$:SeO$_3^{2-}$ = 2:1:1. Optical images show that the as-grown MoSe$_2$ film comprises mainly of continuous monolayers with small portion of voids and thick layers (Figure 4b and 4c). The as-grown MoSe$_2$ film also shows uniform Raman (Figure 4d) and PL spectra (Figure 4e) over the whole wafer. Furthermore, this mixed salts also demonstrate high growth reproducibility (Figure S5a-c). Compared to the growth using Na$_2$MoO$_4$ only with the same Na$^+$:MoO$_4^{2-}$ molar ratio of 2:1 (Figure 2d), the addition of SeO$_3^{2-}$ can effectively improve the growth of MoSe$_2$ monolayers. This further confirms the effectiveness of chalcogen salts in the enhanced CVD growth of 2D TMDCs. It is important to mention that the enhanced MoSe$_2$ growth using (NH$_4$)$_2$MoO$_4$-Na$_2$SeO$_3$ is not caused by the change of transition metal salts, from Na$_2$MoO$_4$ to (NH$_4$)$_2$MoO$_4$. In controlled experiments, there is no MoSe$_2$ observed when only (NH$_4$)$_2$MoO$_4$ is used (Figure S5a and S5b).

Second, it is widely accepted that the Na$^+$ plays vital roles in the growth of TMDCs, which include liquefaction of metal oxides, promoting lateral growth and inducing chalcogen attachment.[17, 20, 23, 42-44] To validate the importance of Na$^+$, the MoSe$_2$ growth is conducted using salt precursors with/without Na$^+$ at fixed SeO$_3^{2-}$ and MoO$_4^{2-}$ molar ratio, i.e. (NH$_4$)$_2$MoO$_4$-Na$_2$SeO$_3$ (16 mM, 1:1) and (NH$_4$)$_2$MoO$_4$-H$_2$SeO$_3$ (16 mM, 1:1). Even with the presence of SeO$_3^{2-}$ in the mixed salts, changing the Na$^+$ to H$^+$ or reducing the Na$^+$ content leads to poor MoSe$_2$ growth (Figure S5c-f and Figure S6). This further confirms the necessity of Na$^+$ being an indispensable element for the growth of MoSe$_2$. A Na$^+$:MoO$_4^{2-}$ molar ratio higher than 2 is needed for the successful growth of continuous MoSe$_2$ film. This is because the Na$^+$ works better than NH$_4^+$ and H$^+$ in stabilizing MoO$_4^{2-}$ during the temperature ramping process. Since the use of $Na_2MoO_4$ only lead to an inferior growth, the results above suggest that both $Na^+$ and $SeO_3^{2-}$ have a synergistic effect on the enhanced CVD growth of large-area wafer-scale $MoSe_2$ film. The effect of different $Na^+:SeO_3^{2-}:MoO_4^{2-}$ molar ratios on the CVD growth of $MoSe_2$ is summarized in Figure 4f.

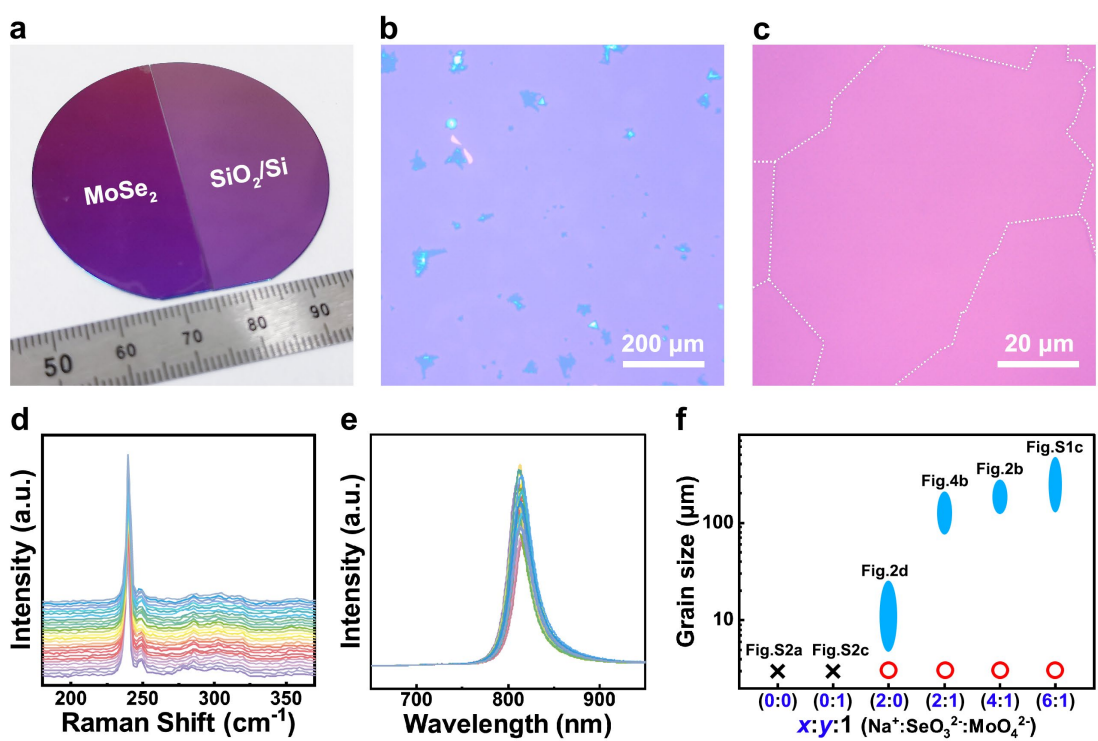

**Figure 4. Contribution of $SeO_3^{2-}$ and $Na^+$ in the $MoSe_2$ growth with mixed salts.** (a) A photo of monolayer $MoSe_2$ film grown on a halved 2-inch $SiO_2$/Si wafer with $(NH_4)_2MoO_4$-$Na_2SeO_3$ (16mM, 1:1). (b, c) Optical images of monolayer $MoSe_2$ film grown with $(NH_4)_2MoO_4$-$Na_2SeO_3$ (16mM, 1:1). The dotted lines indicate the grain boundaries. (d) Raman and (e) PL spectra of the as-grown $MoSe_2$ film. ~ 30 spectra were obtained from the monolayer $MoSe_2$ film shown in (a). (f) The effect of different $Na^+:SeO_3^{2-}:MoO_4^{2-}$ molar ratios on the CVD growth of $MoSe_2$.

Besides the above-mentioned reasons, we also notice that the addition of $Na_2SeO_3$ to $Na_2MoO_4$ results in more uniform precursor coating on $SiO_2/Si$ substrates, even after high-temperature annealing (typical CVD process without introducing Se) (Figure S7). This is particularly important as in our previous study on the $MoS_2$ growth implied that monolayer films formed only on area covered with densely and uniformly distributed, small $Na_2MoO_4$ particles, whereas otherwise on area with low-density, large $Na_2MoO_4$ particles.[18] Note that a high $Na_2SeO_3$: $Na_2MoO_4$ molar ratio of 2 will lead to the aggregation of mixed salt precursors, forming large particles. In this case, only isolated large monolayer $MoSe_2$ flakes are grown (Figure S2c). Hence, careful tuning the molar ratio of the mixed salts is also an important key to the successful growth of continuous monolayer $MoSe_2$ films (Figure 4f).

We further calculated the reaction energies for the possible reactions involved during $MoSe_2$ and $WTe_2$ growth by using the density functional theory (DFT) method. As shown in Table 1, for a typical growth of $MoSe_2$ with $Na_2MoO_4$ only, the reaction energy is -1.1 ev/f.u., Equation (1-1). Increasing the $Na_2SeO_3/Na_2MoO_4$ ratios from 0.5 to 1 and 2, in Equations (1-2) to (1-3) and (1-4), results in a gradual decrease in the reaction energy from -1.96 to -3.00 and -4.93 ev/f.u., respectively. This implies that the addition of $Na_2SeO_3$ is thermodynamically favored for the growth of $MoSe_2$. For the case of $WTe_2$, a similar trend is observed. The introduction of $Na_2TeO_3$ results in a significantly lowered reaction energy from 2.06 ev/f.u. in Equation (2-1) to -0.26 ev/f.u. in Equation (2-2). These results are in consistent to our experimental findings, where the initial poor/limited growth with only the transition metal salts used has improved with the chalcogen salts added, confirming the feasibility of this strategy to the growth of 2D $MoSe_2$ and $WTe_2$.

**Table 1. Possible chemical reactions and corresponding reaction energies for the growth of MoSe$_2$ and WTe$_2$**

| | Possible chemical reaction | Reaction energy (eV/f.u.) |
|---|---|---|
| 1-1 | Na$_2$MoO$_4$ + 3Se + 4H$_2$ = MoSe$_2$ + Na$_2$Se + 4H$_2$O | -1.1 |
| 1-2 | Na$_2$MoO$_4$ + 0.5Na$_2$SeO$_3$ + 3Se + 5.5H$_2$ = MoSe$_2$ + 1.5Na$_2$Se + 5.5H$_2$O | -1.96 |
| 1-3 | Na$_2$MoO$_4$ + Na$_2$SeO$_3$ + 3Se + 7H$_2$ = MoSe$_2$ + 2Na$_2$Se + 7H$_2$O | -3.00 |
| 1-4 | Na$_2$MoO$_4$ + 2Na$_2$SeO$_3$ + 3Se + 10H$_2$ = MoSe$_2$ + 3Na$_2$Se + 10H$_2$O | -4.93 |
| 2-1 | Na$_2$WO$_4$ + 3Te + 4H$_2$ = WTe$_2$ + Na$_2$Te + 4H$_2$O | 2.06 |
| 2-2 | Na$_2$WO$_4$ + 2Na$_2$TeO$_3$ + 3Te + 10H$_2$ = WTe$_2$ + 3Na$_2$Te + 10H$_2$O | -0.26 |

**CONCLUSIONS**

In summary, we demonstrated an enhanced CVD growth of 2D TMD selenides and tellurides using mixed transition metal salts and chalcogen salts. The addition of Na$_2$SeO$_3$ to Na$_2$MoO$_4$ or (NH$_4$)$_2$MoO$_4$, induces a dramatic enhancement in the wafer-scale growth of MoSe$_2$, forming continuous monolayer films with high reproducibility and uniformity at large gain size of 100-250 μm. Similarly, high-efficiency growth of WTe$_2$ is also achieved when mixed salts, Na$_2$WO$_4$-Na$_2$TeO$_3$ are used as growth precursors. This versatile strategy could also be a useful strategy for the growth of other 2D TMDCs and alloys with heterogeneous chalcogen species.

**EXPERIMENTAL SECTION**

**Growth of 2D TMDCs.**

*MoSe$_2$.* First, mixed Na$_2$MoO$_4$-Na$_2$SeO$_3$ solutions with molar ratios of 1:0.5, 1:1 and 1:2 were prepared, respectively. In the mixed salt solutions, the concentration of Na$_2$MoO$_4$ was kept 16 mM. Then, the mixed salt solutions were spin-coated onto SiO$_2$/Si substrates with a speed of 8000 rpm for 30s. The CVD growth was performed in a 2-inch tube furnace. The growth substrates were placed in the center of tube furnace, while a quartz crucible containing ~500 mg Se was placed in the upstream at ~ 300 °C during the growth. 200 sccm Ar/H$_2$ (5% H$_2$) forming gas was introduced as carrier gas. The optimal growth temperature window is 775-850 °C. After CVD growth for 5-10 mins, the tube furnace was cooled down to room temperature in 30 min. To identify the functions of Na (Na$^+$) and Se (SeO$_3^{2-}$) in the enhanced CVD growth of MoSe$_2$, we also preformed controlled CVD growths with Na$_2$MoO$_4$, (NH$_4$)$_2$MoO$_4$, (NH$_4$)$_2$MoO$_4$-Na$_2$SeO$_3$ (1:1) and (NH$_4$)$_2$MoO$_4$-H$_2$SeO$_3$ (1:1), the MoO$_4^{2-}$ is 16 mM in all the aqueous solutions.

*ReSe$_2$.* First, 20 mM mixed NaReO$_4$-Na$_2$SeO$_3$ (1:1) aqueous solution was prepared. The mixed salt deposition and CVD processes are similar to the growth of MoSe$_2$. The optimal growth temperature for ReSe$_2$ is ~750 °C.

*WTe$_2$.* First, 20 mM mixed Na$_2$WO$_4$-Na$_2$TeO$_3$ (1:1) aqueous solution was prepared. Then, the mixed salt solutions were spin-coated onto sapphire substrates with a speed of 5000 rpm for 30s. The growth substrates were placed in the center of the 2-inch tube furnace with a new quartz tube. A quartz crucible containing ~500 mg Te was placed in the upstream at ~ 550 °C during growth. 100 sccm Ar/H$_2$ (5% H$_2$) forming gas was introduced as carrier gas. The optimal growth temperature window is 775-825 °C. After CVD growth for 5 mins, the tube furnace was cooled down to room temperature in 30 min.

**Characterization of 2D TMDCs.**

*Raman and PL.* The micro-Raman/PL was performed using a laser confocal microscope (inVia, Renishaw). The 532-nm excitation laser was focused on the sample surface with a 100× objective lens. Then, Raman/PL signals from the TMDC samples were detected by an electron multiplying CCD detector (Andor) through a grating with 1800 grooves/mm for Raman and 300 grooves/mm for PL. The laser spot size was about 1 μm in diameter. Raman and PL mapping were conducted on the large $MoSe_2$ monolayers with a step of 0.5 μm.

*AFM.* The AFM observations were performed in tapping mode by Asylum MFP-3D origin.

*STEM.* STEM images of $MoSe_2$ monolayer were acquired by using JEOL 2100F microscope equipped with dodecaple correctors and the cold field emission gun operating at 60 kV. The probe current was about 25-30 pA. The convergence semiangle was 35 mrad and the inner acquisition semiangle was 79 mrad.

**Device fabrication and test.**

*Device fabrication.* For the large $MoSe_2$ monolayers transferred on heavily doped silicon substrates with 285-nm-thick $SiO_2$ layer, a layer of photoresist (OFPR-800LB, TOK) was first spin-coated at a speed of 5000 rpm for 30 s. Then, the substrates were baked at 100 °C for 5 mins. A standard LED photolithography process was conducted to define the channel shape (protected by photoresist after development). After developing in NMD-3 (TOK) and rinse with DI water, the surrounding exposed MoSe$_2$ monolayers were etched with oxygen plasma. Next, a second LED photolithography was conducted to define the electrode patterns. Tri-layer of Cr/Pd/Au (1/10/50 nm) was deposited as contacts via e-beam deposition. The final MoSe$_2$-FETs have channel length of ~5 μm and with of ~20 μm (Figure S3). Similar fabrication process was also conducted on the transferred ReSe$_2$ and WTe$_2$ without the etching procedure. For ReS$_2$- and WTe$_2$-FETs, bi-layer of Cr/Au (1/50 nm) were deposited as contacts.

*Measurement of TMDC-based FETs.* To test the transport properties of MoSe$_2$-FETs, the devices were loaded in a vacuum chamber with a pressure of ~2×10$^{-3}$ Pa. The backgate bias ($V_{gs}$) was scanned forward from -10 V to 50 V and backward from 50 V to -10 V with a step of 1 V, the source-drain bias ($V_{ds}$) is 2 V. For the transport curves plotted in Figure 2j, the y-axis was normalized to conductivity using the channel current and device dimensions. For the WTe$_2$-FETs, the backgate bias ($V_{gs}$) was scanned forward from -30 V to 30 V and backward from 30 V to -30 V with a step of 1 V, the source-drain bias ($V_{ds}$) is 1 V.

**Density functional theory (DFT) calculation.**

The reaction energies were evaluated using the DFT method within the generalized gradient approximation of the Perdew, Burke, and Ernzernhof functional as implemented in the Vienna ab initio simulation package.[45, 46] Electron−ion interactions were described using projector-augmented wave pseudopotentials.[47] A plane-wave kinetic-energy cutoff of 600 eV and a k-spacing of 0.2 Å$^{-1}$ in reciprocal space were used to ensure that the energy converged to better than 1 meV/atom. All of the crystal structures were obtained from the Materials Project database.[48]

**ASSOCIATED CONTENT**

Supporting Information.

This material is available free of charge via the Internet at http://pubs.acs.org.

The Supporting Information include: (i) schematic illustration of mixed-salt CVD process, (ii) characterization of as-grown 2D $MoSe_2$ and $ReSe_2$, (iii) characterization of $MoSe_2$-FETs, (iv) calculated energy of chemicals.

**AUTHOR INFORMATION**

Corresponding Authors

S. Li: li.shisheng@nims.go.jp

T. Taniguchi: taniguchi.takaaki@nims.go.jp

ORCID

Shisheng Li: 0000-0001-9301-5559

Yung-Chang Lin: 0000-0002-3968-7239

Jinhua Hong: 0000-0002-6406-1780

Bo Gao: 0000-0003-1183-656X

Hong En Lim: 0000-0003-0347-8897

Xu Yang: 0000-0001-8195-5850

Song Liu: 0000-0003-3390-7795

Yoshitaka Tateyama: 0000-0002-5532-6134

Yoshiki Sakuma: 0000-0001-68047217

Kazuhito Tsukagoshi: 0000-0001-9710-2692

Kazu Suenaga: 0000-0002-6107-1123

Takaaki Taniguchi: 0000-0002-8460-5431

Notes

The authors declare no competing financial interests.

**ACKNOWLEDGMENT**

S.L. acknowledges the support from JSPS-KAKENHI (19K15399, 21K04839). T.T. acknowledges the support from JSPS-KAKENHI (17K19187). Y.-C.L. and K.S. acknowledge support from the JST-CREST (JPMJCR20B1, JPMJCR20B5, JPMJCR1993), JSPS-KAKENHI (18K14119), JSPS A3 Foresight Program, and Kazato Research Encouragement Prize. Y.S. acknowledges the support from JSPS-KAKENHI (17H03241). H.E.L. acknowledges the support from JSPS-KAKENHI (19K15393, 21K14498). B.G. and Y.T. thank the support by MEXT as "Program for Promoting Researches on the Supercomputer Fugaku (Fugaku Battery & Fuel Cell Project), Grant JPMXP1020200301, and the supercomputer at NIMS. S.L. acknowledges all staff members of the Nanofabrication Group at NIMS for their support.

**References**

1. Fiori, G.; Bonaccorso, F.; Iannaccone, G.; Palacios, T.; Neumaier, D.; Seabaugh, A.; Banerjee, S. K.; Colombo, L., Electronics Based on Two-Dimensional Materials. *Nat. Nanotechnol.* **2014**, *9*, 768-779.

2. Gong, C.; Zhang, Y.; Chen, W.; Chu, J.; Lei, T.; Pu, J.; Dai, L.; Wu, C.; Cheng, Y.; Zhai, T., Electronic and Optoelectronic Applications Based on 2D Novel Anisotropic Yransition Metal Dichalcogenides. *Adv. Sci.* **2017**, *4*, 1700231.

3. Su, S.-K.; Chuu, C.-P.; Li, M.-Y.; Cheng, C.-C.; Wong, H.-S. P.; Li, L.-J., Layered Semiconducting 2D Materials for Future Transistor Applications. *Small Structures*, **2021**, *2*, 2000103.

4. Shi, J.; Huan, Y.; Zhao, X.; Yang, P.; Hong, M.; Xie, C.; Pennycook, S.; Zhang, Y., Two-Dimensional Metallic Vanadium Ditelluride as a High-Performance Electrode Material. *ACS Nano* **2021**, *15*, 1858-1868.

5. Zhang, Z.; Gong, Y.; Zou, X.; Liu, P.; Yang, P.; Shi, J.; Zhao, L.; Zhang, Q.; Gu, L.; Zhang, Y., Epitaxial Growth of Two-Dimensional Metal–Semiconductor Transition-Metal

Dichalcogenide Vertical Stacks (VSe$_2$/MX$_2$) and Their Band Alignments. *ACS Nano* **2018**, *13*, 885-893.

6. Mleczko, M. J.; Xu, R. L.; Okabe, K.; Kuo, H.-H.; Fisher, I. R.; Wong, H.-S. P.; Nishi, Y.; Pop, E., High Current Density and Low Thermal Conductivity of Atomically Thin Semimetallic WTe$_2$. *ACS Nano* **2016**, *10*, 7507-7514.

7. Lee, C.-S.; Oh, S. J.; Heo, H.; Seo, S.-Y.; Kim, J.; Kim, Y. H.; Kim, D.; Ngome Okello, O. F.; Shin, H.; Sung, J. H., Epitaxial van der Waals Contacts Between Transition-Metal Dichalcogenide Monolayer Polymorphs. *Nano Lett.* **2019**, *19*, 1814-1820.

8. Tan, J.; Li, S.; Liu, B.; Cheng, H.-M., Structure, Preparation, and Applications of 2D Material‐Based Metal–Semiconductor Heterostructures. *Small Structures* **2021**, *2*, 2000093.

9. Kang, K.; Xie, S.; Huang, L.; Han, Y.; Huang, P. Y.; Mak, K. F.; Kim, C.-J.; Muller, D.; Park, J., High-Mobility Three-Atom-Thick Semiconducting Films with Wafer-Scale Homogeneity. *Nature* **2015**, *520*, 656-660.

10. Wang, D.; Luo, F.; Lu, M.; Xie, X.; Huang, L.; Huang, W., Chemical Vapor Transport Reactions for Synthesizing Layered Materials and Their 2D Counterparts. *Small* **2019**, *15*, 1804404.

11. Lee, Y. H.; Zhang, X. Q.; Zhang, W.; Chang, M. T.; Lin, C. T.; Chang, K. D.; Yu, Y. C.; Wang, J. T. W.; Chang, C. S.; Li, L. J., Synthesis of Large‐Area MoS$_2$ Atomic Layers with Chemical Vapor Deposition. *Adv. Mater.* **2012**, *24*, 2320-2325.

12. Shi, Y.; Li, H.; Li, L.-J., Recent Advances in Controlled Synthesis of Two-Dimensional Transition Metal Dichalcogenides via Vapour Deposition Techniques. *Chem. Soc. Rev.* **2015**, *44*, 2744-2756.

13. Huang, J.-K.; Pu, J.; Hsu, C.-L.; Chiu, M.-H.; Juang, Z.-Y.; Chang, Y.-H.; Chang, W.-H.; Iwasa, Y.; Takenobu, T.; Li, L.-J., Large-Area Synthesis of Highly Crystalline WSe$_2$ Monolayers and Device Applications. *ACS Nano* **2014**, *8*, 923-930.

14. Li, S.; Wang, S.; Tang, D.-M.; Zhao, W.; Xu, H.; Chu, L.; Bando, Y.; Golberg, D.; Eda, G., Halide-Assisted Atmospheric Pressure Growth of Large WSe$_2$ and WS$_2$ Monolayer Crystals. *Appl. Mater. Today* **2015**, *1*, 60-66.

15. Zhou, J.; Lin, J.; Huang, X.; Zhou, Y.; Chen, Y.; Xia, J.; Wang, H.; Xie, Y.; Yu, H.; Lei, J., A Library of Atomically Thin Metal Chalcogenides. *Nature* **2018**, *556*, 355-359.

16. Liu, C.; Xu, X.; Qiu, L.; Wu, M.; Qiao, R.; Wang, L.; Wang, J.; Niu, J.; Liang, J.; Zhou, X., Kinetic Modulation of Graphene Growth by Fluorine Through Spatially Confined Decomposition of Metal Fluorides. *Nat. Chem.* **2019**, *11*, 730-736.

17. Li, S.; Lin, Y.-C.; Zhao, W.; Wu, J.; Wang, Z.; Hu, Z.; Shen, Y.; Tang, D.-M.; Wang, J.; Zhang, Q., Vapour–Liquid–Solid Growth of Monolayer $MoS_2$ Nanoribbons. *Nat. Mater.* **2018**, *17*, 535-542.

18. Li, S.; Lin, Y.-C.; Liu, X.-Y.; Hu, Z.; Wu, J.; Nakajima, H.; Liu, S.; Okazaki, T.; Chen, W.; Minari, T., Wafer-Scale and Deterministic Patterned Growth of Monolayer $MoS_2$ via Vapor–Liquid–Solid Method. *Nanoscale* **2019**, *11*, 16122-16129.

19. Liu, H.; Qi, G.; Tang, C.; Chen, M.; Chen, Y.; Shu, Z.; Xiang, H.; Jin, Y.; Wang, S.; Li, H., Growth of Large-Area Homogeneous Monolayer Transition-Metal Disulfides via a Molten Liquid Intermediate Process. *ACS Appl. Mater. & Interfaces* **2020**, *12*, 13174-13181.

20. Chang, M.-C.; Ho, P.-H.; Tseng, M.-F.; Lin, F.-Y.; Hou, C.-H.; Lin, I.-K.; Wang, H.; Huang, P.-P.; Chiang, C.-H.; Yang, Y.-C., Fast Growth of Large-Grain and Continuous $MoS_2$ Films Through a Self-Capping Vapor-Liquid-Solid Method. *Nat. Commun.* **2020**, *11*, 3682.

21. Zuo, Y.; Yu, W.; Liu, C.; Cheng, X.; Qiao, R.; Liang, J.; Zhou, X.; Wang, J.; Wu, M.; Zhao, Y., Optical Fibres with Embedded Two-Dimensional Materials for Ultrahigh Nonlinearity. *Nat. Nanotechnol.* **2020**, *15*, 987-991.

22. Li, S.; Hong, J.; Gao, B.; Lin, Y. C.; Lim, H. E.; Lu, X.; Wu, J.; Liu, S.; Tateyama, Y.; Sakuma, Y., Tunable Doping of Rhenium and Vanadium into Transition Metal Dichalcogenides for Two-Dimensional Electronics. *Adv. Sci.* **2021**, *8*, 2004438.

23. Cai, Z.; Lai, Y.; Zhao, S.; Zhang, R.; Tan, J.; Feng, S.; Zou, J.; Tang, L.; Lin, J.; Liu, B., Dissolution-Precipitation Growth of Uniform and Clean Two Dimensional Transition Metal Dichalcogenides. *Nat. Sci. Rev.* **2021**, *8*, nwaa115.

24. Gong, Y.; Lin, Z.; Ye, G.; Shi, G.; Feng, S.; Lei, Y.; Elias, A. L.; Perea-Lopez, N.; Vajtai, R.; Terrones, H., Tellurium-Assisted Low-Temperature Synthesis of $MoS_2$ and $WS_2$ Monolayers. *ACS Nano* **2015**, *9*, 11658-11666.

25. Cui, F.; Wang, C.; Li, X.; Wang, G.; Liu, K.; Yang, Z.; Feng, Q.; Liang, X.; Zhang, Z.; Liu, S., Tellurium-Assisted Epitaxial Growth of Large-area, Highly Crystalline ReS$_2$ Atomic Layers on Mica Substrate. *Adv. Mater.* **2016**, *28*, 5019-5024.

26. Tonndorf, P.; Schmidt, R.; Böttger, P.; Zhang, X.; Börner, J.; Liebig, A.; Albrecht, M.; Kloc, C.; Gordan, O.; Zahn, D. R., Photoluminescence Emission and Raman Response of Monolayer MoS$_2$, MoSe$_2$, and WSe$_2$. *Opt. Express* **2013**, *21*, 4908-4916.

27. Liu, Z.; Amani, M.; Najmaei, S.; Xu, Q.; Zou, X.; Zhou, W.; Yu, T.; Qiu, C.; Birdwell, A. G.; Crowne, F. J., Strain and Structure Heterogeneity in MoS$_2$ Atomic Layers Grown by Chemical Vapour Deposition. *Nat. Commun.* **2014**, *5*, 5246.

28. Yang, S.; Chen, Y.; Jiang, C., Strain Engineering of Two-Dimensional Materials: Methods, Properties, and Applications. *InfoMat* **2021**, *3*, 397-420.

29. Wang, X.; Gong, Y.; Shi, G.; Chow, W. L.; Keyshar, K.; Ye, G.; Vajtai, R.; Lou, J.; Liu, Z.; Ringe, E., Chemical Vapor Deposition Growth of Crystalline Monolayer MoSe$_2$. *ACS Nano* **2014**, *8*, 5125-5131.

30. Han, H.-V.; Lu, A.-Y.; Lu, L.-S.; Huang, J.-K.; Li, H.; Hsu, C.-L.; Lin, Y.-C.; Chiu, M.-H.; Suenaga, K.; Chu, C.-W., Photoluminescence Enhancement and Structure Repairing of Monolayer MoSe$_2$ by Hydrohalic Acid Treatment. *ACS Nano* **2016**, *10*, 1454-1461.

31. Chen, M.-W.; Ovchinnikov, D.; Lazar, S.; Pizzochero, M.; Whitwick, M. B.; Surrente, A.; Baranowski, M.; Sanchez, O. L.; Gillet, P.; Plochocka, P., Highly Oriented Atomically Thin Ambipolar MoSe$_2$ Grown by Molecular Beam Epitaxy. *ACS Nano* **2017**, *11*, 6355-6361.

32. Meng, Y.; Ling, C.; Xin, R.; Wang, P.; Song, Y.; Bu, H.; Gao, S.; Wang, X.; Song, F.; Wang, J., Repairing Atomic Vacancies in Single-Layer MoSe$_2$ Field-Effect Transistor and Its Defect Dynamics. *npj Quantum Mater.* **2017**, *2*, 16.

33. Li, Y.; Zhang, K.; Wang, F.; Feng, Y.; Li, Y.; Han, Y.; Tang, D.; Zhang, B., Scalable Synthesis of Highly Crystalline MoSe$_2$ and Its Ambipolar Behavior. *ACS Appl. Mater. & Interfaces* **2017**, *9*, 36009-36016.

34. Chen, X.; Hu, P.; Song, K.; Wang, X.; Zuo, C.; Yang, R.; Wang, J., CVD Growth of Large–Scale Hexagon-Like Shaped MoSe$_2$ Monolayers with Sawtooth Edge. *Chem. Phys. Lett.* **2019**, *733*, 136663.

35. Somvanshi, D.; Ber, E.; Bailey, C. S.; Pop, E.; Yalon, E., Improved Current Density and Contact Resistance in Bilayer MoSe$_2$ Field Effect Transistors by AlO$_x$ Capping. *ACS Appl. Mater. & Interfaces* **2020**, *12*, 36355-36361.

36. Kim, M.; Seo, J.; Kim, J.; Moon, J. S.; Lee, J.; Kim, J.-H.; Kang, J.; Park, H., High-Crystalline Monolayer Transition Metal Dichalcogenides Films for Wafer-Scale Electronics. *ACS Nano* **2021**, *15*, 3038-3046.

37. Chen, Z.; Liu, H.; Chen, X.; Chu, G.; Chu, S.; Zhang, H., Wafer-Size and Single-Crystal MoSe$_2$ Atomically Thin Films Grown on GaN Dubstrate for Light Emission and Harvesting. *ACS Appl. Mater. Interfaces* **2016**, *8*, 20267-20273.

38. Li, J.; Yan, W.; Lv, Y.; Leng, J.; Zhang, D.; Coileáin, C. Ó.; Cullen, C. P.; Stimpel-Lindner, T.; Duesberg, G. S.; Cho, J., Sub-Millimeter Size High Mobility Single Crystal MoSe$_2$ Monolayers Synthesized by NaCl-Assisted Chemical Vapor Deposition. *RSC Adv.* **2020**, *10*, 1580-1587.

39. Li, N.; Wang, Q.; Shen, C.; Wei, Z.; Yu, H.; Zhao, J.; Lu, X.; Wang, G.; He, C.; Xie, L., Large-Scale Flexible and Transparent Electronics Based on Monolayer Molybdenum Disulfide Field-Effect Transistors. *Nat. Electron.* **2020**, *3*, 711-717.

40. Zhou, Z.; Xu, T.; Zhang, C.; Li, S.; Xu, J.; Sun, L.; Gao, L., Enhancing Stability by Tuning Element Ratio in 2D Transition Metal Chalcogenides. *Nano Res.* **2021**, *14*, 1704-1710.

41. Li, J.; Cheng, S.; Liu, Z.; Zhang, W.; Chang, H., Centimeter-Scale, Large-Area, Few-Layer 1T′-WTe$_2$ Films by Chemical Vapor Deposition and Its Long-Term Stability in Ambient Condition. *J. Phys. Chem. C* **2018**, *122*, 7005-7012.

42. Yang, P.; Zou, X.; Zhang, Z.; Hong, M.; Shi, J.; Chen, S.; Shu, J.; Zhao, L.; Jiang, S.; Zhou, X., Batch Production of 6-Inch Uniform Monolayer Molybdenum Disulfide Catalyzed by Sodium in Glass. *Nat. Commun.* **2018**, *9*, 979.

43. Xie, C.; Yang, P.; Huan, Y.; Cui, F.; Zhang, Y., Roles of Salts in the Chemical Vapor Deposition Synthesis of Two-Dimensional Transition Metal Chalcogenides. *Dalton Trans.* **2020**, *49*, 10319-10327.

44. Kim, H.; Ovchinnikov, D.; Deiana, D.; Unuchek, D.; Kis, A., Suppressing Nucleation in Metal–Organic Chemical Vapor Deposition of MoS$_2$ Monolayers by Alkali Metal Halides. *Nano Lett*. **2017**, *17*, 5056-5063.

45. Perdew, J. P.; Burke, K.; Ernzerhof, M., Generalized gradient approximation made simple. *Phys. Rev. Lett.* **1996**, *77*, 3865.

46. Kresse, G.; Joubert, D., From ultrasoft pseudopotentials to the projector augmented-wave method. *Phy. Rev. B* **1999**, *59*, 1758.

47. Blöchl, P. E., Projector augmented-wave method. *Phy. Rev. B* **1994**, *50*, 17953.

48. Jain, A.; Ong, S. P.; Hautier, G.; Chen, W.; Richards, W. D.; Dacek, S.; Cholia, S.; Gunter, D.; Skinner, D.; Ceder, G., Commentary: The Materials Project: A materials genome approach to accelerating materials innovation. *APL Mater.* **2013**, *1*, 011002.

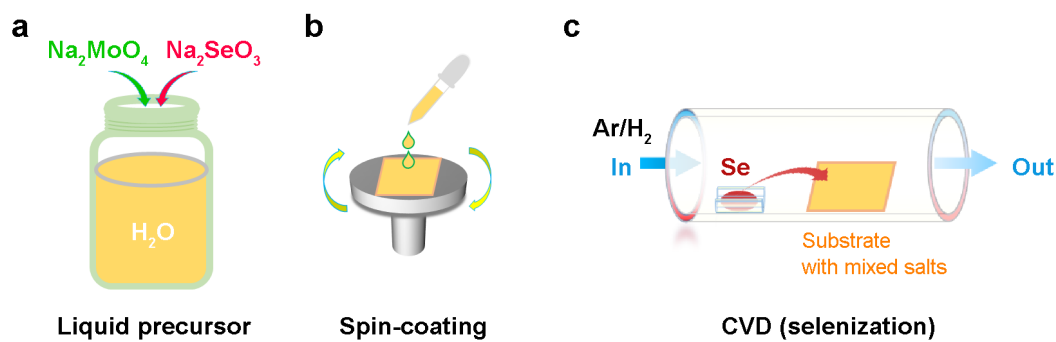

**Figure S1.** Schematic illustrations of the enhanced CVD growth of monolayer MoSe$_2$ films using mixed Na$_2$MoO$_4$/Na$_2$SeO$_3$. (a) The preparation of liquid precursors by dissolving Na$_2$MoO$_4$ and Na$_2$SeO$_3$ in DI H$_2$O. (b) Depositing mixed salts on growth substrate via spin-coating of the liquid precursor. (c) CVD growth of MoSe$_2$ films via selenization of the mixed Na$_2$MoO$_4$/Na$_2$SeO$_3$ on substrate.

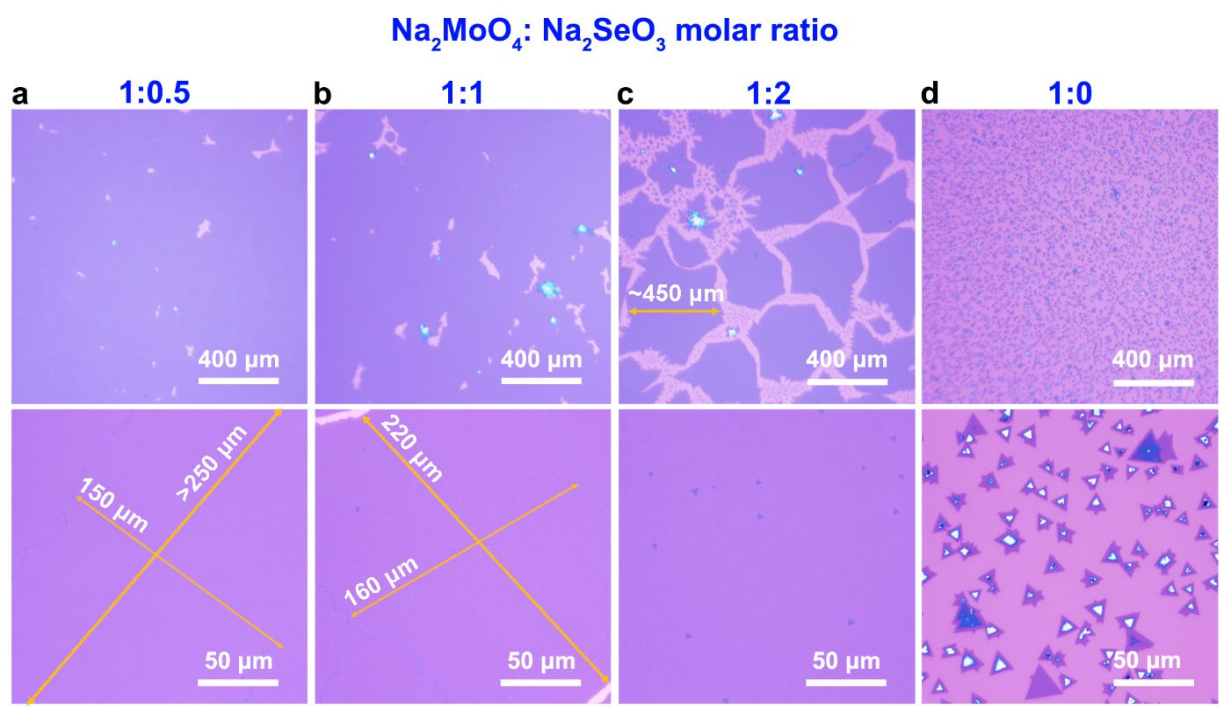

**Figure S2. Optical images of MoSe₂ grown on SiO₂/Si substrates with different Na₂MoO₄:Na₂SeO₃ ratios.** (a) 1:0.5, (b) 1:1, (c) 1:2 and (d) 1:0. Large MoSe₂ monolayers with grain size in the range of 150 - 450 µm can be grown when using mixed Na₂MoO₄-Na₂SeO₃ as growth precursors.

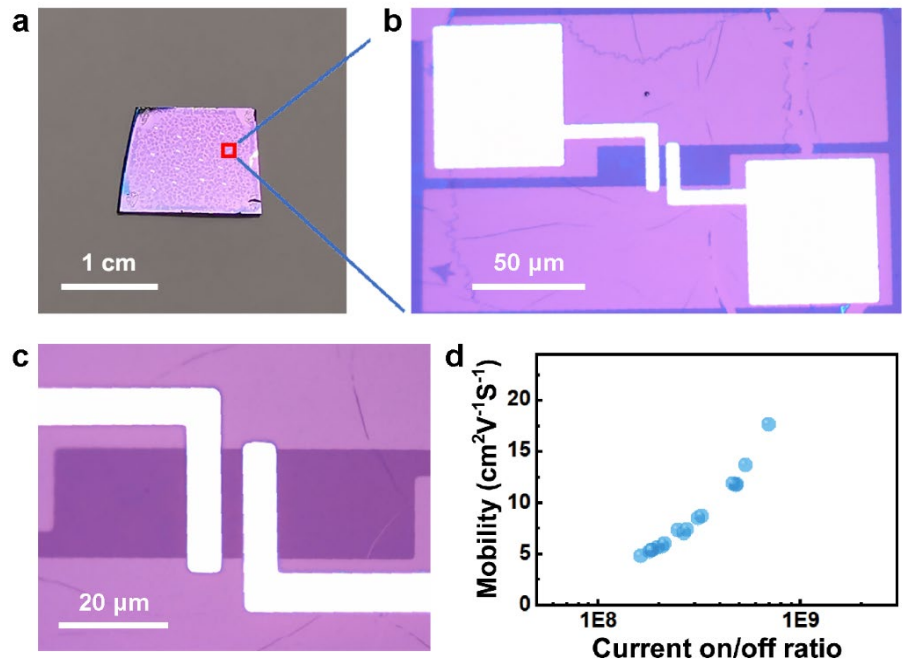

**Figure S3. Monolayer MoSe$_2$-FETs.** (a) An array of monolayer MoSe$_2$-FETs fabricated with MoSe$_2$ film transferred on 285 nm SiO$_2$ (285 nm)/Si substrate (~10×10 mm$^2$). (b, c) Optical images of a single MoSe$_2$-FET. Photolithography and oxygen plasma were employed to define the channel shape and etch surrounding MoSe$_2$. (d) Statistical result of the electron mobility vs. current on/off ratio for a total of 16 MoSe$_2$-FETs measured.

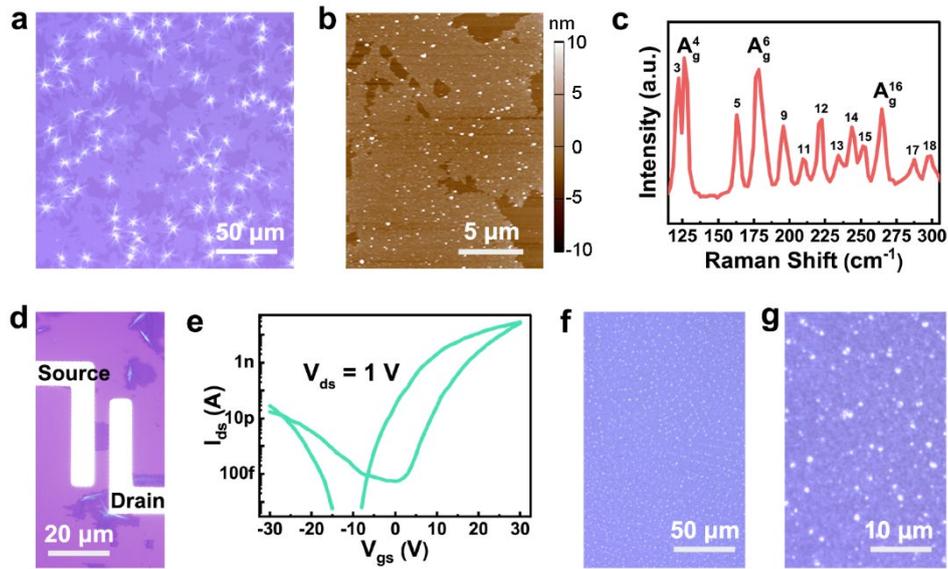

**Figure S4. Characterization of ReSe$_2$ grown with mixed NaReO$_4$-Na$_2$SeO$_3$.** (a) An optical image showing the as-grown ReSe$_2$ flakes on sapphire substrate. (b) An AFM image of the ReSe$_2$ flakes transferred onto a SiO$_2$/Si substrate. (c) Typical Raman spectrum of the as-grown ReSe$_2$ flakes. (d, e) Typical (d) optical image and (e) transport curve of a ReSe$_2$-FET. (f, g) Optical images of as-grown small ReSe$_2$ flakes using NaReO$_4$ only.

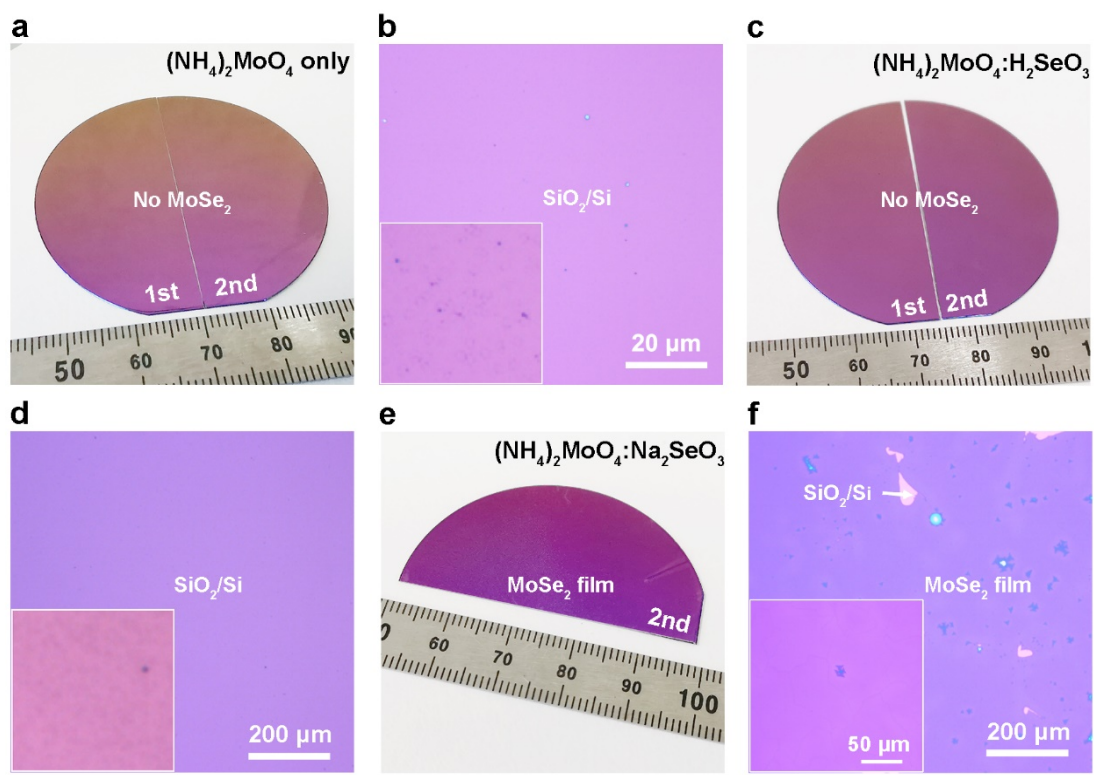

**Figure S5. Controlled experiments using different salt precursors.** (a, b) A photo and optical images show the repeated growth of MoSe$_2$ with (NH$_4$)$_2$MoO$_4$ only (No Na$^+$ and SeO$_3^{2-}$), and there is no MoSe$_2$ was observed after the CVD growth. Inset shows an area of 10 ×10 μm$^2$. (c, d) A photo and optical images show the repeated growth of MoSe$_2$ with (NH4)$_2$MoO$_4$-H$_2$SeO$_3$ (No Na$^+$), and there is no MoSe$_2$ was able to grow the mixed salts. Inset shows an area of 10 ×10 μm$^2$. (e, f) A photo and optical images show the second growth of monolayer MoSe$_2$ film with (NH$_4$)$_2$MoO$_4$-Na$_2$SeO$_3$ (16 mM, 1:1) on a half of 2-inch SiO$_2$/Si wafer, and (f) optical images show the morphology of as-grown monolayer MoSe$_2$ film.

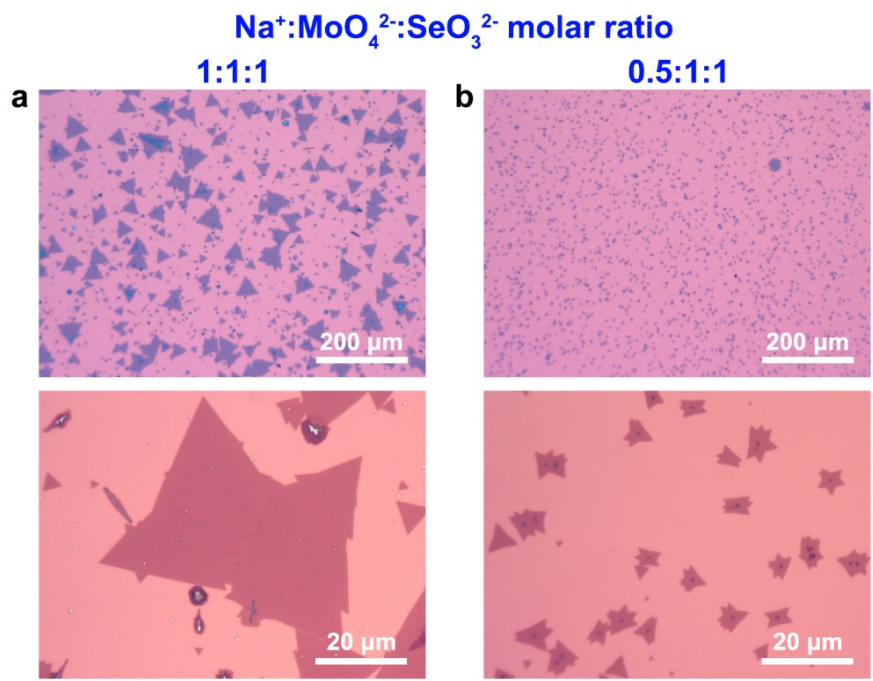

**Figure S6. Controlled experiments using different Na$^+$ contents.** Optical images of MoSe$_2$ monolayers grown with Na$^+$:MoO$_4^{2-}$:SeO$_3^{2-}$ molar ratios of (a) 1:1:1 and (b) 0.5:1:1. The molar concentration of MoO$_4^{2-}$ and SeO$_3^{2-}$ is 16 mM, prepared using a mixture of Na$_2$MoO$_4$, (NH$_4$)$_2$MoO$_4$ and H$_2$SeO$_3$.

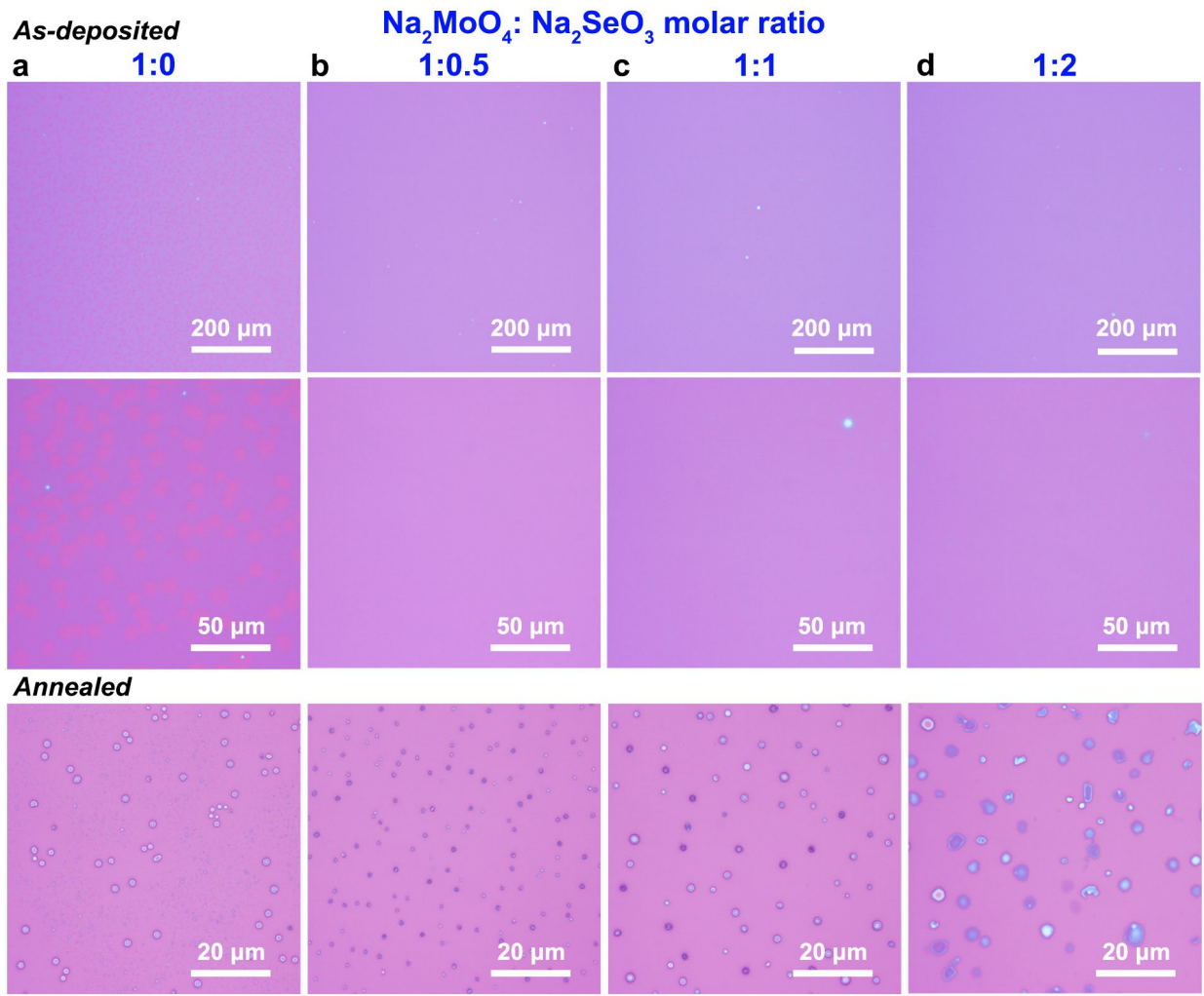

**Figure S7. Comparison of spin-coated mixed-salt precursors with different Na₂MoO₄:Na₂SeO₃ molar ratios.** The Na₂MoO₄:Na₂SeO₃ molar ratios are (a) 1:0; (b) 1:0.5; (c) 1:1 and (d) 1:2. In all salt precursors, the Na₂MoO₄ is kept as 16 mM. The spin-coating condition is 8000 rpm for 30 s. The first two rows show the morphologies of as-prepared salt precursors on SiO₂/Si substrates. The third row shows the size and area-density of salt precursors after annealing in a typical CVD process without introducing selenium.

**Table S1. Calculated energy of chemicals**

|  | Energy (eV/f.u.) |  | Energy (eV/f.u.) |
|---|---|---|---|
| $MoSe_2$ | -19.973 | $WTe_2$ | -19.575 |
| $Na_2MoO_4$ | -47.597 | $Na_2WO_4$ | -50.638 |
| $Na_2SeO_3$ | -29.870 | $Na_2TeO_3$ | -29.831 |
| $Na_2Se$ | -9.387 | $Na_2Te$ | -8.607 |
| Se | -3.506 | Te | -3.142 |
| $H_2$ | -6.773 | $H_2O$ | -14.229 |